\documentclass{article}

\usepackage{arxiv}

\usepackage[utf8]{inputenc} 
\usepackage[T1]{fontenc}    
\usepackage{hyperref}       
\usepackage{url}            
\usepackage{booktabs}       
\usepackage{amsfonts}       
\usepackage{nicefrac}       
\usepackage{microtype}      
\usepackage{lipsum}
\usepackage{graphicx}

\usepackage{pgf-pie}
\usepackage{stfloats}

\usepackage[table]{xcolor}
\definecolor{azulUC3M}{RGB}{0,0,102}
\definecolor{gray97}{gray}{.97}
\definecolor{gray75}{gray}{.75}
\definecolor{gray45}{gray}{.45}

\usepackage[a-1b]{pdfx}

\usepackage{hyperref}
\hypersetup{colorlinks=true,
	linkcolor=black, 
	urlcolor=blue} 

\usepackage{amsmath,amssymb,amsfonts,amsthm}

\usepackage{txfonts} 
\usepackage[T1]{fontenc}
\usepackage[utf8]{inputenc}

\usepackage[english]{babel} 
\usepackage[babel, english=american]{csquotes}
\AtBeginEnvironment{quote}{\small}

\usepackage{fancyhdr}
\fancypagestyle{plain}{\pagestyle{fancy}}

\usepackage{multirow} 
\usepackage{caption} 
\DeclareCaptionOption{compatibility}{false}
\usepackage{floatrow} 
\usepackage{array} 
\newcolumntype{P}[1]{>{\centering\arraybackslash}p{#1}}
\DeclareCaptionFormat{upper}{#1#2\uppercase{#3}\par}
\usepackage{graphicx}
\graphicspath{{imagenes/}} 

\usepackage{chngcntr} 

\usepackage{listings}

\lstdefinestyle{estilo}{ frame=Ltb,
	framerule=0pt,
	aboveskip=0.5cm,
	framextopmargin=3pt,
	framexbottommargin=3pt,
	framexleftmargin=0.4cm,
	framesep=0pt,
	rulesep=.4pt,
	backgroundcolor=\color{gray97},
	rulesepcolor=\color{black},
	basicstyle=\ttfamily\footnotesize,
	keywordstyle=\bfseries,
	stringstyle=\ttfamily,
	showstringspaces = false,
	commentstyle=\color{gray45},     
	numbers=left,
	numbersep=15pt,
	numberstyle=\tiny,
	numberfirstline = false,
	breaklines=true,
	xleftmargin=\parindent
}

\lstdefinelanguage{json}{
  basicstyle=\ttfamily,
  numbers=left,
  numberstyle=\tiny,
  stepnumber=1,
  numbersep=5pt,
  showstringspaces=false,
  breaklines=true,
  frame=single
}

\begin{document}

\title{Design and Development of an Intelligent LLM-based LDAP Honeypot}

\author{Javier Jiménez Román
\\Universidad Carlos III de Madrid\\
Avda de la Universidad 30\\
Legan\'es, Madrid, Spain\\
\And
Florina Almenares-Mendoza \\
Universidad Carlos III de Madrid\\
Avda de la Universidad 30\\
Legan\'es, Madrid, Spain\\
\texttt{florina@it.uc3m.es},
\And
Alfonso S\'anchez-Maci\'an \\
Universidad Carlos III de Madrid\\
Avda de la Universidad 30\\
Legan\'es, Madrid, Spain\\
\texttt{alfonsan@it.uc3m.es},
}

\maketitle

\begin{abstract}
Cybersecurity threats continue to increase, with a growing number of previously unknown attacks each year targeting both large corporations and smaller entities. This scenario demands the implementation of advanced security measures, not only to mitigate damage but also to anticipate emerging attack trends. In this context, deception tools have become a key strategy, enabling the detection, deterrence, and deception of potential attackers while facilitating the collection of information about their tactics and methods. Among these tools, honeypots have proven their value, although they have traditionally been limited by rigidity and configuration complexity, hindering their adaptability to dynamic scenarios. The rise of artificial intelligence, and particularly general-purpose Large Language Models (LLMs), is driving the development of new deception solutions capable of offering greater adaptability and ease of use. This work proposes the design and implementation of an LLM-based honeypot to simulate an LDAP server, a critical protocol present in most organizations due to its central role in identity and access management. The proposed solution aims to provide a flexible and realistic tool capable of convincingly interacting with attackers, thereby contributing to early detection and threat analysis while enhancing the defensive capabilities of infrastructures against intrusions targeting this service.
\end{abstract}

\keywords{
Deception Technology, AI-powered Honeypot, Large Language Models (LLM), LDAP Protocol, Cyber Threat Intelligence, Threat Detection
}

\section{Introduction}

Lightweight Directory Access Protocol (LDAP) is the standard protocol for 
querying and modifying information stored in directory services. These 
directories are widely used for identity and credential management (e.g., 
Microsoft Active Directory, OpenLDAP, or enterprise SSO systems) and 
typically contain highly sensitive information such as usernames, email 
addresses, roles, and passwords. Such information is extremely valuable to 
attackers, as it enables the use of valid credentials, lateral movement, 
privilege escalation, and even provides internal reconnaissance of an 
organization’s structure \cite{redhat_ldap_auth}.

The LanDscAPe study analyzed the global LDAP exposure surface and identified 
more than 82,000 publicly reachable LDAP servers \cite{LanDscAPe}. 
The results revealed that a significant portion of the LDAP ecosystem is 
misconfigured, allowing unauthorized access to sensitive data. Among the 
most severe issues were the presence of servers still running the deprecated 
LDAP v2 protocol and instances exposed without any authentication (enabling 
anonymous access). Other problems included systems remaining vulnerable 
despite the availability of patches and around 34\% of TLS enabled servers 
relying on weak or outdated cryptographic configurations.

While these findings reflect the risks of publicly exposed servers, it is 
likely that similar misconfigurations may also affect internal corporate 
environments, extending the threat surface to organizational infrastructures.

A paradigmatic example of LDAP being exploited indirectly is the Log4Shell 
vulnerability (CVE-2021-44228) \cite{cve2021-44228}. Although the flaw 
originated in the Log4j library, attackers leveraged LDAP servers
as a channel to deliver malicious payloads via JNDI. 
This case demonstrates that, even when not the root cause of a 
vulnerability, LDAP can still play a critical role in large-scale 
exploitation campaigns, highlighting the importance of securing LDAP within 
a broader security context.

A direct impact on LDAP itself can be observed in one of its most common 
vulnerabilities called LDAP Injection. This type of attack is similar to SQL 
Injection and occurs when applications dynamically construct LDAP queries 
from user input without proper sanitization. As a result, attackers may 
manipulate the query to gain unauthorized access to information or even 
alter the behavior of the directory service \cite{owasp_ldap_injection_prevention}.

Given the critical role of LDAP in identity management and the 
vulnerabilities that affect it, it becomes essential to explore defensive 
mechanisms capable of detecting and analyzing malicious activity. Honeypots 
emerge as a suitable approach, as a security mechanism designed to simulate 
vulnerable resources, such as operating systems, protocols, or networks, 
with two main purposes.

On the one hand, honeypots serve as a research and intelligence gathering tool, attracting 
attackers and recording their activities. The subsequent analysis of these interactions provides 
valuable insights into adversarial behavior and their TTPs (Tactics, Techniques, and Procedures). 
This type of information constitutes threat intelligence that can be integrated into other 
security tools to improve detection capabilities and prevent future incidents. In this way, 
honeypots can contribute to the identification of new IOCs from emerging campaigns and even enable 
the discovery of zero-day vulnerabilities \cite{greynoise_zero_day,bleepingcomputer_juniper_rce}.

On the other hand, honeypots can also play a defensive and reactive role by acting as decoys 
within an organization’s infrastructure. In this approach, they lure attackers and redirect 
malicious activity toward non-critical systems, thereby reducing risk to production assets. At the 
same time, they enable the detection of active threats and provide actionable intelligence to 
reinforce defenses in real time \cite{bleepingcomputer_fake_ddos}.

Different types of honeypots exist, and the most common classification is based on the level of 
interaction allowed with the attacker \cite{Kambow2014HoneypotsT}. Low-interaction 
honeypots emulate services or protocols only in a very basic manner (e.g., SSH, FTP, MySQL, HTTP) 
without fully implementing their functionality. Their purpose is mainly to detect connection 
attempts (such as collecting IP addresses) without allowing arbitrary code execution. Their main 
advantages are ease of deployment and low risk of compromise, although they provide limited 
intelligence.

At the next level, medium-interaction honeypots provide more realistic services and allow limited 
interaction with the attacker within a controlled environment. This enables the collection of more 
detailed information, such as exploitation attempts or recorded command sequences, while 
maintaining moderate risk by preventing full system compromise.

Finally, high-interaction honeypots are the most advanced and realistic but also the riskiest. 
They typically involve full systems, often virtual machines or containers with intentionally 
vulnerable versions exposing them to real attacks. In these environments, adversaries may fully 
compromise the system and deploy malware, offering an opportunity to study TTPs in depth. From a 
forensic and threat discovery perspective, these honeypots are extremely valuable, though they 
require strict isolation to avoid their misuse for lateral movement or incorporation into botnets.

An alternative classification, independent of interaction level, encompasses other honeypot 
modalities. One example is honeytokens, which are not full systems but lightweight digital decoys 
such as fake credentials, configuration files, beaconed documents, dummy API keys, or database 
records that should never be accessed legitimately. Interaction with these artifacts triggers an 
alert, signaling intrusion or potential data exfiltration. A practical example is the EDR 
SentinelOne, which places specific files in the “C:\textbackslash” path so that they appear first when sorted 
alphabetically. If the system detects that one of these files begins to be encrypted, it 
interprets the action as a potential ransomware attempt \cite{sentinelone_honeytokens}.

Another modality is honeynets, which, instead of simulating a single service or system, reproduce 
entire networks of interconnected honeypots to emulate full enterprise or industrial 
infrastructures. This approach enables the study of sophisticated attacks, observation of lateral 
movement techniques, and analysis of how adversaries coordinate their actions within a realistic 
environment.

Finally, email/web honeypots are decoys specifically designed to attract spam or phishing 
campaigns. They are implemented through fake email addresses or controlled websites that collect 
information about attackers, the infrastructure they use, and their distribution tactics. These 
mechanisms help improve spam filters, detect active campaigns, and gather intelligence on 
malicious actors \cite{bleepingcomputer_fake_azure_honeypots}.

Building on these concepts, this work focuses on the design of a honeypot 
specifically adapted to the LDAP protocol, leveraging recent advances in 
Large Language Models. The goal is to combine the strengths of honeypots as 
deception tools with the adaptability of LLMs, thereby addressing existing 
gaps in the literature. 

The scope of this project is to demonstrate that an LLM-based honeypot can 
simulate LDAP with reasonable accuracy and robust responses, covering a 
broad range of the operations defined by the protocol. In addition, the work 
contributes a dataset of LDAP traffic, addressing the current lack of 
publicly available sources. The tool has also been designed to systematically 
record logs of all interactions, enabling their subsequent analysis and the 
extraction of threat intelligence. Since the objective is to develop a tool from 
scratch, the focus has been placed on building a foundation that 
delivers essential functionality while maintaining the flexibility to be 
extended in future iterations. Features beyond the scope of this work, 
such as TLS encryption of LDAP traffic, are not included, nor is the aim to 
replicate the full complexity of an Active Directory environment, which 
encompasses additional services and characteristics beyond the protocol 
itself. Likewise, the purpose of this work is not to perform an in-depth optimization of 
the underlying LLM, but rather to validate its viability as the core of a functional 
honeypot system.

It should also be noted that the project was developed without access to 
large-scale dedicated infrastructure for training and deploying LLMs. 
Instead, it relied on cloud-based resources, such as Google Colab Pro with 
an NVIDIA A100 GPU, which are powerful but still subject to practical 
constraints in availability and configuration. This setup was sufficient for 
the objectives of the project, emphasizing feasibility and reproducibility 
over exhaustive optimization

The remainder of this work is structured as follows. Section II provides the background and related work, starting with an overview of traditional honeypots, followed by early applications of artificial intelligence in deception systems, and concluding with recent approaches based on LLMs. Section III details the design and implementation of the system, including the dataset generation process, the overall architecture, and the description of each module that composes the honeypot. Section IV reports the evaluation methodology and experimental 
results, introducing a custom set of metrics specifically designed to assess the behavior of an LLM-based LDAP honeypot. Finally, Section V summarizes the main contributions and outlining possible directions for future work.

\section{Background \& Related Work}

Before presenting the design and implementation of the proposed honeypot, it is essential to review 
the current state of the art in deception technologies and related approaches. This section provides 
the necessary background to contextualize the work, starting with an overview of traditional 
honeypots, followed by early applications of artificial intelligence in deception systems, and 
concluding with recent efforts that integrate LLMs. This review highlights the evolution of the 
field, the limitations of existing approaches, and the opportunities that motivate the development 
of an LLM-based honeypot for the LDAP protocol.

\subsection{Traditional Honeypots}

The concept of a honeypot in cybersecurity dates back to the late 1980s and early 1990s, with 
documented cases such as those described by Clifford Stoll (1989) and Bill Cheswick (1991) 
\cite{wikipedia_cuckoos_egg, wikipedia_honeypot}.  Since then, honeypots have evolved from experimental solutions to 
widely adopted implementations in research environments, corporate networks, government agencies, and law enforcement, 
adapting to emerging technologies and an increasingly complex threat landscape. In thiscontext, the Honeynet Project 
was established in 1999 as a non-profit international initiative “dedicated to the research and development of 
honeypot and honeynet technologies with the goal of improving information security through the collection and analysis 
of real attack data” \cite{honeynet_project}. The project remains highly active today, hosting events and publishing 
valuable resources, while maintaining an extensive collection of honeypots covering various technologies, along with 
repositories useful for threat intelligence.

At present, there is a wide range of active, fully open-source projects that are regularly updated,
many of them curated and promoted by the Honeynet Project. Among these, T-Pot \cite{tpotce_github} stands out as an 
open-source platform that integrates multiple types of honeypots and complementary tools into a single, easily 
deployable environment. Similar to the Honeynet Project, T-Pot consolidates various threat collection and analysis 
technologies, providing centralized data gathering alongside monitoring, visualization, and correlation capabilities. 
Related to this ecosystem, GreedyBear \cite{greedybear_github} focuses on aggregating and 
correlating data from multiple honeypots, and offers a set of public feeds 
\cite{greedybear_honeynet} for automated consumption and threat intelligence analysis. Other
notable projects, although older and no longer actively maintained, include Honeyd 
\cite{honeyd_project} and Honeytrap \cite{honeytrap_github}, both of which were 
highly relevant in their time: Honeyd for its ability to flexibly and cost-effectively emulate 
multiple systems and services on a single physical machine, and Honeytrap for its modular design 
enabling malicious traffic capture and analysis across different protocols and services.

When focusing specifically on honeypots designed for the LDAP protocol, available developments are 
considerably more limited. One notable example is a study conducted by researchers from a Danish 
university, aimed at analyzing various attacks targeting LDAP by deploying honeypots to log 
malicious activity \cite{9799363}. However, the collected data was
not made publicly available, and despite attempts to contact the authors, it has not been possible 
to incorporate this traffic into the dataset used in this work. There are also other attempts with a 
certain level of maturity in implementing LDAP within honeypots, such as the work by Dimas \& 
Charles \cite{10276516}. In this study, the authors employed the 
previously mentioned Honeytrap framework to analyze four services: Ethereum, SMTP, FTP, and LDAP. In 
the case of LDAP, only a very small number of samples were collected, most of which contained a few 
basic LDAP operations. According to the researchers, these appeared to be primarily attempts to 
determine whether the service was active, suggesting a simple port scan targeting the exposed 
service. Notably, this work reflects on how honeypots can facilitate the capture of real-world 
attacks, particularly against less monitored services such as LDAP. 

More recently, a low-interaction honeypot project for LDAP has been published, which focuses exclusively on recording authentication  attempts by capturing credentials \cite{ldap_honeypot_github}. This project does not implement other 
LDAP operations and lacks LDAPS support. Another relevant repository is Chameleon \cite{chameleon_github}, which implements modules for multiple protocols and lists LDAP as future work; however, it has not been updated for two years. The lack of open-source projects focused on LDAP may be attributed to the wide range of operations defined by 
the protocol and the inherent complexity of realistically simulating complete interactions.

\subsection{Early Applications of AI in Honeypots}

Before the exponential growth of artificial intelligence driven by the Transformers architecture,
advances in Natural Language Processing (NLP), and the emergence of general-purpose models such as
ChatGPT, several works had already been published that applied AI algorithms to bring new approaches
to the field of deceptive tools such as honeypots. An example is QRASSH \cite{8484261}, which applies Deep Q-Learning 
with a neural network as an approximator to control a self-adaptive SSH honeypot. Its goal is to combine honeypots with
reinforcement learning to select the best action for each attacker behavior, maximizing the utility
of the interaction. The reward function is designed to incentivize adversary behaviors, such as malware downloads, 
that enrich the logs and facilitate the extraction of more relevant Indicators of
Compromise (IOCs).

Another example is the work by Huang \& Zhu, which uses a Semi-Markov model to maximize the
attacker’s engagement time in a honeynet while minimizing the risk of compromising the decoy through
adaptive network policies \cite{Huang_2019}.

In the field of Industrial Control Systems (ICS) protocols, a notable case is NeuralPot
\cite{9219712}, which trains two Deep Neural Network (GAN and Autoencoder) models with Modbus traffic
captures, achieving realistic responses to Modbus queries and actively misleading intruders.

\subsection{Honeypots Powered by LLM-Based AI}

With the rapid emergence of artificial intelligence and the widespread adoption of 
LLMs — general-purpose models that have begun to permeate every field — various 
implementations have also been proposed in the domain of honeypots. A 2024 study explored this 
concept by conducting a series of experiments on datasets simulating SSH servers, using GPT-3.5 
without fine-tuning, to evaluate whether the use of LLMs could enhance honeypot capabilities \cite{10703029}.

The results highlighted several key findings: (1) simple commands yielded better performance compared to more complex 
ones; (2) maintaining conversational context across sessions proved to be important; (3) the authors concluded that 
LLMs are particularly well-suited for request–response protocols; (4) compared to high-interaction honeypots that 
emulate protocols, LLM-based systems reduce the likelihood of a real system compromise; and (5) certain limitations 
were identified, such as response latency — which could alert an attacker — and the difficulty for LLMs to provide 
accurate real-time date and time information when required.

Similar ideas are observed in other studies, such as \cite{10295397}, which also leverage GPT-based models to simulate 
an SSH environment. In addition, this work focuses on securing the model itself, aiming to mitigate direct attacks 
against the LLM, such as prompt injection.

The application of LLMs within deceptive tools is not limited to honeypots. Several studies have 
explored this combination to generate honeytokens including passwords, configuration files, and 
other decoys using automated prompt engineering techniques \cite{reti2024acthoneytokengeneratorinvestigation}. These 
experiments have revealed that not all LLMs respond with the same level of effectiveness, indicating that model 
selection plays a significant role in the reliability and quality of the generated artifacts.

It can be observed that this is a field in which numerous works have been developed in recent years, including some 
notable examples integrated into broader projects previously mentioned, such as T-Pot \cite{tpotce_github}. T-Pot also 
incorporates advanced LLM-based honeypots, including Beelzebub \cite{beelzebub_honeypot}, which simulates a wide range 
of protocols, and Galah \cite{galah_github}, which focuses specifically on the HTTP protocol.

From the reviewed literature, two main approaches can be identified when integrating LLMs into 
honeypot systems. The first involves using LLMs through APIs provided by vendors such as OpenAI or 
Google, typically combined with prompt engineering techniques to refine responses. The second 
consists of fine-tuning open-source models to achieve behavior more closely aligned with the 
specific system or protocol being simulated.

\begin{itemize}
\item \textbf{Fine-tuning}: This approach involves retraining an open-source LLM on a domain-specific dataset so that 
its responses are tailored to the exact context, system, or protocol it aims to emulate. Several works fall into this 
category, including a honeypot simulating the SSH protocol \cite{Otal_2024}, others targeting the 
MySQL database protocol \cite{10607309}, more domain-specific protocols such as those used in the industrial sector \cite{vasilatos2025llmpotdynamicallyconfiguredllmbased}, and even honeypots designed to mimic a real-world API \cite{SEZGIN2025104458}.

\item \textbf{API-based}: Studies in this group rely on accessing LLMs via external APIs to simulate interactive 
environments without the need to host or train the model locally. This approach offers the ability to leverage state-of-the-art proprietary models. Examples include the simulation of SSH consoles \cite{Sladi__2024,10808265} and various shell environments \cite{wang2025honeygptbreakingtrilemmaterminal,HoneyLLM,10735663}, where LLMs are tasked with convincingly emulating user–system interactions in real time.
\end{itemize}

\section{Design \& Implementation}

Following the literature review, this work aims to broaden the existing range of solutions and 
address identified gaps, supported by evidence from prior studies showing that delegating a 
honeypot’s responses to an LLM can yield significant improvements. Among the most frequently cited 
advantages, adaptability stands out, enabling the system to respond dynamically to varied attacker 
behaviors. Additionally, by not exposing a real system, the risk of an actual machine compromise is inherently 
reduced. However, this does not entirely eliminate security concerns, as LLMs themselves introduce a distinct set of 
risks, as outlined in the OWASP LLM Top 10 \cite{owasp_llm_top10}.

Probably the most significant drawback of an LLM-based honeypot is response 
latency, which is influenced by the number of tokens the model must generate 
and the hardware on which it is executed. These factors can be adjusted to 
reduce response time, either by sacrificing some degree of accuracy or by 
deploying the model on high-performance infrastructure.

Within the broad spectrum of potential systems, protocols, and functionalities that could be 
simulated, LDAP was selected as the target for this study. This decision is motivated by the 
protocol’s critical role in modern infrastructures and the sensitivity of the information it 
manages. Furthermore, this work addresses the absence of any publicly documented LLM-based honeypot capable of 
simulating the LDAP protocol, as no prior publications evidencing such a system have been identified.

As the project evolved, several key design decisions were made. One of these concerned whether to 
implement the system through an API-based approach or to apply some form of fine-tuning. Ultimately, the latter option 
was chosen, as it was believed to offer better results and enable the development of a more comprehensive tool. To 
support this decision, the works reviewed in the preceding section were analyzed, leading to a set of conclusions for 
each approach, which are summarized in Table~\ref{tab:llm_api_vs_ft}.

\renewcommand{\arraystretch}{1.5}
\begin{table}[htbp]
\caption{LLM: API vs. Fine-tuning}
\begin{center}
\footnotesize
\begin{tabular}{|m{3cm}|m{2.5cm}|m{3cm}|}
\hline
\textbf{Aspect} & \textbf{API} & \textbf{Fine-tune} \\
\hline
Implementation Ease & High & Low \\
\hline
Control & Limited by prompts & Full output control \\
\hline
Deploy Time & Days–weeks & Weeks–months \\
\hline
Realism & Good if base strong & Higher; domain-trained \\
\hline
Cost & Low–medium & High \\
\hline
\end{tabular}
\label{tab:llm_api_vs_ft}
\end{center}
\end{table}
\renewcommand{\arraystretch}{1.0}

During the development phase, few-shot learning was tested as a potential approach. This technique ranges from zero-shot (no examples provided) to n-shot (a limited set of examples), guiding a model to perform a task without full retraining. Initial experiments with models such as LLaMA showed that it could handle very simple scenarios; however, as soon as the complexity increased — particularly due to LDAP’s variability in operations, attributes, and structures — the model failed to provide adequate responses. The limited examples were insufficient for the model to generalize effectively, reinforcing the conclusion that fine-tuning was the most appropriate solution for the problem.

Another relevant aspect of the implementation is the adjustment of data handling to ensure that both input and output are in JSON format. The model receives a request in JSON and returns its response in the same structure. This choice is driven by several factors: JSON offers a convenient and human-readable format for programming, development, and testing; it enables the serialization of the LLM’s textual output, allowing the necessary fields to be reconstructed in an orderly manner for the client’s response; and it is widely used in logging, with native support in many tools for importing and processing JSON data.

Some of the most popular tools used within organizational ecosystems that can integrate this 
type of data include Azure Monitor, Splunk, and the ELK/Logstash stack. Azure Monitor, 
Microsoft’s monitoring service, can ingest JSON-formatted files through data collection rules 
\cite{microsoft_azure_monitor_json}. Splunk provides a dedicated sourcetype for processing JSON 
logs \cite{somerford_json_splunk}. Finally, the ELK stack leverages Logstash, a plugin capable 
of parsing JSON messages and decomposing them into structured fields 
\cite{coralogix_logstash_tutorial}.

\subsection{Dataset} 

Regarding dataset generation, this step was necessary to perform fine-tuning, as no public 
datasets containing LDAP traffic were found. To address this, three LDIF (LDAP Data Interchange 
Format) files were created using the tool LDIF Generator \cite{ldifgen_github}, which allows 
the construction of complex directory structures by customizing parameters such as the number 
of users, groups, or the depth of nested OUs. For each file, a different domain was selected: 
domain.example.org, corp.local, and globex.internal. These structures were then further 
enriched to increase diversity and enable additional operations. Specifically, attributes such 
as email, phone number, department number, job title/position, and passwords were added, the 
latter making it possible to perform authentication operations \cite{faker_docs}. This 
enrichment process was partially automated through a custom script, which ensured consistency 
across generated entries while maintaining variability. 
Once defined, the files were imported one by one into an OpenLDAP server.

\begin{figure}[htbp]
\centering
\begin{tikzpicture}
\pie[
    text=legend,
    radius=2,
    sum=auto,
    color={blue!40, red!40, green!40, orange!40}
]{
43.12/queries (43.12\%),
40.67/authentications (40.67\%),
8.87/others (8.87\%),
7.34/modifications (7.34\%)
}
\end{tikzpicture}
\caption{Distribution of LDAP Operations in the Dataset.}
\label{fig:ldap_ops_openldap_simplified}
\end{figure}
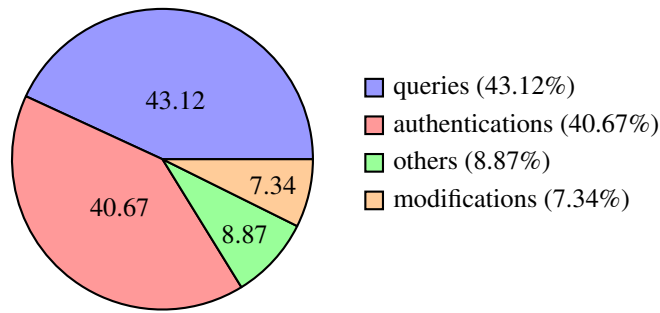

For each OpenLDAP configuration, a plan was designed and executed that comprised a broad range of 
operations, including authentications (40.67\%), queries (43.12\%), creations (1.83\%), 
modifications (7.34\%), moves (0.31\%), deletions (1.53\%), comparisons (4.28\%) and extended 
operations (0.92\%), the distribution can be observed in Fig.~\ref{fig:ldap_ops_openldap_simplified}. 
In total, the dataset contains 328 records and also includes failed operations.
The percentages reflect a diverse distribution of operation types, with queries and 
authentications dominating, which is consistent with typical LDAP usage in production 
environments: “A typical client session consists of one or more bind operations, followed by a 
number of search operations, and then an unbind.” 
\cite{4447466}.

To perform these, most commands were executed with the \textit{ldap-utils} package, complemented by actions from Nmap scripts, Metasploit modules, and other graphical tools. These tools were included because it is possible that attackers may rely on them during reconnaissance phases in real-world scenarios.

\begin{figure}[htbp]
  \centering
  \includegraphics[width=180pt]{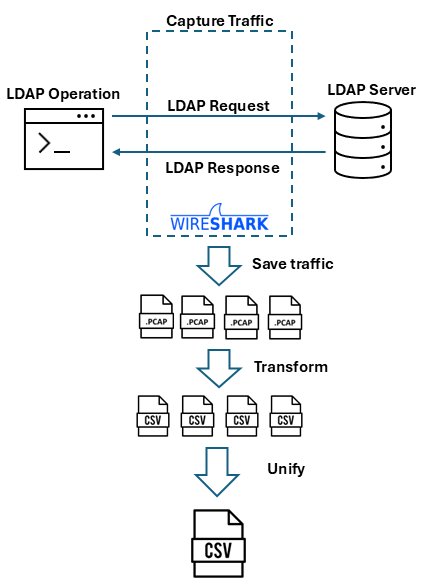}
  \caption{Dataset Creation Process}
  \label{fig:dataset_creation}
\end{figure}

In order to construct the dataset, each operation defined in the plan was executed while capturing the corresponding network traffic with Wireshark, storing every action individually in .pcap, .pcapng, or .cap files. Once the LDAP traffic was collected, a custom parser script was used to automate its transformation into a single CSV file. The process decoded LDAP messages from the raw captures, preserved the request–response mapping based on \textit{messageID}, and 
exported them as structured JSON pairs, as can be observed in Fig.~\ref{fig:ldap_log_example}. The resulting unified CSV contains all operations in two columns —input and output— representing the interaction between client and server. This process is summarized in Fig.~\ref{fig:dataset_creation}.

\begin{figure}[htbp]
\centering
\begin{minipage}{0.48\textwidth}
\begin{lstlisting}[language=json, 
                   basicstyle=\ttfamily\scriptsize, 
                   frame=single, 
                   numbers=none, 
                   backgroundcolor=\color{white}]
{
  "messageID": 9,
  "protocolOp": {
    "searchRequest": {
      "baseObject": "dc=corp,dc=local",
      "scope": 2,
      "sizeLimit": 20,
      "filter": {
        "equalityMatch": {
          "attributeDesc": "objectClass",
          "assertionValue": "person"
        }
      },
      "attributes": ["cn","sn","mail","objectClass"]
    }
  }
}
\end{lstlisting}
\end{minipage}
\hfill
\begin{minipage}{0.48\textwidth}
\begin{lstlisting}[language=json, 
                   basicstyle=\ttfamily\scriptsize, 
                   frame=single, 
                   numbers=none, 
                   backgroundcolor=\color{white}]
{
  "messageID": 9,
  "protocolOp": {
    "searchResEntry": {
      "objectName": "cn=prokosch oosterink,ou=Accounting,...",
      "attributes": [
        {"type": "objectClass",
         "vals": ["top","person","organizational\\Person","inetOrgPerson"]},
        {"type": "cn", "vals": ["prokosch oosterink"]},
        {"type": "sn", "vals": ["oosterink"]},
        {"type": "mail", "vals": ["new.admin@corp.local"]}
      ]
    }
  }
}
\end{lstlisting}
\end{minipage}
\caption{Example of LDAP request and response encoded in JSON.}
\label{fig:ldap_log_example}
\end{figure}

It is important to note that a single request can generate multiple responses, particularly in 
search operations where multiple entries may be returned, followed by a final message indicating the end of the response. As this characteristic is inherent to LDAP, the dataset was explicitly designed to capture it, ensuring that each request can be associated with several nested JSON responses.

\subsection{Architecture}

The architecture developed in this work is composed of several coordinated components that 
together simulate the behavior of an LLM-based honeypot for the LDAP protocol. The workflow 
begins with the reception of LDAP requests by a local listener, which parses them and 
transforms them into JSON format to be interpreted by the model. These requests are then 
forwarded to a remote service running on Colab, where the LLM generates the corresponding 
responses.

The responses are transmitted back in JSON format, reconstructed into ASN.1/BER, and finally 
returned to the LDAP client as if they originated from a real server. At this stage, all interactions are also recorded in dedicated logs, ensuring that the traffic generated by the honeypot can later be analyzed to extract valuable insights into attacker behavior and potential threats.

\begin{figure*}[t]
  \centering
  \includegraphics[width=\textwidth]{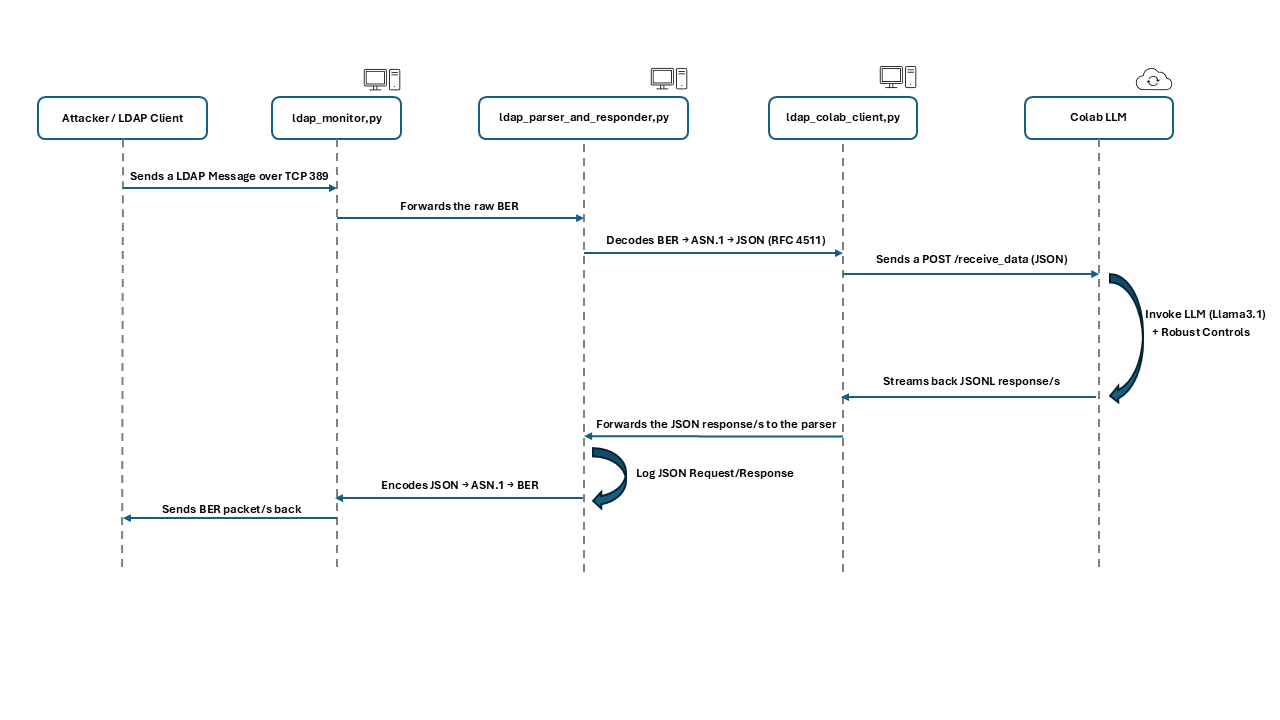}
  \caption{Overview of the Honeypot-LLM LDAP system architecture}
  \label{fig:architecture}
\end{figure*}

The Fig.~\ref{fig:architecture} provides a schematic overview of this workflow, which is subsequently 
explained in detail through the description of each individual module.

\subsection{Listener LDAP} 

The \textit{ldap\_monitor.py} script serves as the socket server, responsible for receiving network frames and handling incoming raw LDAP requests encoded in BER. It converts the LDAP data into hexadecimal format and delegates the parsing and response generation to other components. Once the responses are produced, it also manages the construction of the corresponding TCP packets. If an unbind operation is received from the client, it is ignored. This script acts as the entry point of the entire system, and the IP address and port on which it runs can be configured as needed.

\subsection{LDAP Orchestrator} 

LDAP as a protocol follows the ASN.1 (Abstract Syntax Notation One) standard, which defines how data structures are 
represented so that clients and servers share a common syntax to exchange messages. 
Each operation is subject to specific rules defined in RFC 4511 \cite{rfc4511}. LDAP traffic is not transmitted in plain text but rather encoded using BER (Basic Encoding Rules), which serialize ASN.1 messages into bytes before transmission. Consequently, in order for the honeypot to process incoming messages and generate responses that comply with the RFC specification, a Python script \textit{ldap\_parser\_and\_responder.py} was designed to translate a BER-encoded hex stream containing LDAP data into a plain-text JSON format, in accordance with RFC 4511.

A widely used library, \textit{pyasn1} \cite{pyasn1_github}, implements the ASN.1 specification along with its 
primary encoding rules (BER, DER, CER). Building on this foundation, additional libraries such as \textit{pyasn1-modules} 
\cite{pyasn1_modules_github} extend its functionality by providing Python classes for a variety of RFCs. However, none of these modules specifically implement RFC 4511. An alternative project, \textit{asn1ate} \cite{asn1ate_github}, offers the capability to transform an ASN.1 definition into a Python class, thereby enabling encoding and decoding in compliance with a given RFC.

At the beginning of this work, several experiments were conducted using asn1ate to generate Python classes compliant with RFC 4511. While the resulting implementation was able to successfully handle basic operations, such as authentication and simple searches, it failed when dealing with more complex requests. Eventually, the open-source library \textit{pyasn1-ldap} \cite{pyasn1_ldap_pypi} was identified. This library, which also leverages \textit{asn1ate}, provides a functional Python implementation of LDAP message parsing. Since it demonstrated robust handling of various LDAP operations, it was adopted as the foundation for this project.

Once the request is parsed, it is forwarded to the LLM, delegating this action to the next 
script in the chain. When the response from the LLM is received, either as a single JSON object 
or as multiple ones, it is reconstructed one by one into ASN.1 and re-encoded into BER before 
being returned to the listener.

In addition to generating and transmitting responses, the system incorporates a dedicated 
logging mechanism. Every request–response pair is stored in JSON format, including metadata 
such as the exact timestamp of the interaction and the client’s IP address. This feature not 
only ensures traceability of all simulated traffic but also provides a structured dataset that 
can later be leveraged for threat intelligence, forensic analysis, and system debugging. 
Moreover, the logs can be integrated into the workflow of real-time security mechanisms, 
enabling the automatic triggering of alerts when suspicious activity is detected.

\subsection{Bridge LLM}

The \textit{ldap\_colab\_client.py} script functions as an HTTP bridge between the local orchestrator and the inference service deployed in Colab. Specifically, it sends the LDAP request already serialized to JSON by the orchestrator via a POST request to the public "/receive\_data" endpoint exposed through ngrok. The client supports response streaming, enabling line by line processing in cases where the LLM produces multiple JSON objects, as commonly occurs with search operations. In this way, each response generated by the LLM in Colab can be retrieved individually.

Alternative approaches (e.g., SSH tunnels or VPNs) were considered for connectivity, but an HTTP client using ngrok was selected due to its ease of integration and because the expected workload did not exceed the free tier quotas. Within this architecture, the ngrok token is configured on the server side (the Flask service running in Colab) to open the tunnel and publish a public URL. The client itself does not transmit credentials; instead, it simply posts the request to the exposed URL (e.g., https://…ngrok-free.app/receive\_data).

\subsection{Deployment Platforms}

Due to the lack of dedicated infrastructure to execute an LLM, the Google Colab platform was used, 
as it provides access to powerful GPUs capable of performing fine-tuning within a reasonable time 
frame and executing the model efficiently. Alongside this platform, Unsloth 
\cite{unsloth_notebooks_docs}, an open-source project that provides tools 
to simplify the fine-tuning of LLMs, was employed. Unsloth enables faster and more efficient 
training by reducing resource consumption without compromising output quality. As a result, even 
large-scale LLMs can be adapted within modest computational environments. In this work, Unsloth 
significantly facilitated the fine-tuning process through its well-structured Colab notebook 
templates, which provide guided introductions and initial benchmarking for experimentation.

Another noteworthy feature integrated into these notebooks is LoRA (Low-Rank Adaptation) 
\cite{hu2021loralowrankadaptationlarge}, a fine-tuning technique designed to adapt large-scale models without retraining all of their original parameters. Instead of updating the entire network, LoRA introduces low-rank matrices into specific layers, dramatically reducing the number of parameters that need to be optimized. This approach results in significantly lower resource consumption while maintaining the ability to fine-tune state-of-the-art models.

Among the available options, the LLaMA 3.1 (8B) model was selected for this work, primarily due to its popularity, efficiency, and extensive community support. Furthermore, it includes native 
function-calling capabilities, which were initially considered useful for validating the correctness of JSON outputs during response generation, although this feature was ultimately not utilized in practice.

Our model is fine-tuned using a sequence-to-sequence framework with 6 training epochs. 
The batch size is set to 16 per device, with no gradient accumulation. We employ a 
cosine learning rate scheduler with a base learning rate of $5 * 10^{-5}$ and a dynamic warm-
up ratio of 0.12. Gradient clipping is applied with a maximum norm of 0.5, and weight 
decay is set to 0.01. Optimization is performed using AdamW with 8-bit precision. For 
numerical stability, mixed precision is enabled with BF16, which is natively supported 
on NVIDIA A100 GPUs. The training process is conducted with PyTorch and the TRL library.

The choice of a cosine learning rate scheduler is motivated by its ability to smoothly decrease the learning rate, 
which helps avoid abrupt changes during optimization and often results in better convergence compared to step-based 
schedules \cite{loshchilov2017sgdrstochasticgradientdescent}. Gradient clipping with a maximum norm of 0.5 is applied to stabilize training, as it prevents exploding gradients in transformer architectures, particularly when using large batch sizes or long sequences \cite{pascanu2013difficultytrainingrecurrentneural}. Additionally, a weight decay of 0.01 is used to regularize the model, reducing overfitting by penalizing large parameter values, a technique widely adopted in fine-tuning large language models \cite{loshchilov2019decoupledweightdecayregularization}

During the training phase in the Google Colab environment, several relevant features were incorporated. For each request–response pair in the dataset, a prompt was applied together with specific instructions. The primary objective was to guide the LLM not only toward predicting a valid LDAP response but, more importantly, toward generating a correctly structured JSON output. Particular attention was paid to the proper use of quotation marks, braces, and brackets to ensure 
strict compliance with the required syntax.

In the prediction phase, two prompt modes were defined for the LLM. The first, automatic mode, allowed the model to infer context autonomously from the request itself, adapting the response according to attributes present in the input, for example the domain specified or the email domain included in a search filter. In this way, if the domain contained a “.es” TLD, the model tended to 
produce names and values in Spanish in its responses.

The second mode, manual, was designed for the operator to explicitly configure a fixed context. In 
this case, parameters such as the base distinguished name (DN), language, and organizational type to 
be simulated could be specified, ensuring greater coherence with the operational environment in 
which the tool was deployed.

In both modes, a set of common instructions was enforced: replicating the same \textit{messageID} from the 
request within the response, ensuring the syntactic validity of the JSON, and consistently reusing 
the input DN throughout the response.

Additional safeguards were implemented in the Python script executed in Colab. These included an estimated calculation of the number of tokens required, based on an analysis of the data generated during dataset construction. Similar to the strategy proposed by Ragsdale and Boppana (2023) \cite{10295397}, careful prompt engineering was applied to minimize randomness and reduce unnecessary token consumption. Accordingly, a maximum token allocation was set depending on the operation type. For instance, broad search queries, or requests with a higher sizeLimit, were assigned a large number of maximum tokens. While simple authentication operations (\textit{bindRequest}) required much shorter outputs and thus significantly fewer tokens.

If the operation type was not recognized, the function default value is set to 1,000 tokens, providing a sufficient margin to attempt handling unexpected outputs.

Finally, due to the token limitations of the LLM, it was possible for responses to exceed the estimated maximum and become truncated. To mitigate this issue, the system evaluated the generated output by splitting it into JSON blocks and discarding any incomplete ones. In the specific case of search operations, if the response was detected to be truncated or missing the appropriate closing \textit{searchResDone}, the system automatically generated a closing JSON object including the same 
\textit{messageID}, thereby ensuring communication consistency and structural validity.

\section{Evaluation \& Results}

\subsection{Custom Evaluation Framework}

In the context of this work, traditional evaluation metrics commonly applied to machine learning models are not directly suitable for assessing the behavior of an LLM-based honeypot simulating LDAP. Accuracy, understood as the proportion of outputs identical to the expected ones, would be ineffective since exact string-level matches are rarely achievable: distinguished names (DNs) may vary in the order of their attributes, optional fields such as \textit{diagnosticMessage} may legitimately differ, and the ordering of \textit{searchResEntry} results in a search operation is not semantically relevant. Likewise, metrics such as Precision, Recall, or F1-score are designed for tasks where 
there is a well-defined set of expected and predicted elements (e.g., classification or entity detection). Applying them here would require a rigorous definition of what constitutes a “correct” response: whether it is the presence of all required attributes, the use of the correct \textit{messageID}, or the consistent return of a \textit{searchResDone} message. For these reasons, a custom evaluation system was designed, tailored to the specific characteristics of LDAP and the objectives of this honeypot. That system consists of different key properties of the LDAP structure, which are objectively measurable without penalizing natural variation.

\begin{itemize}

\item \textbf{Syntax Pass Rate}: This metric measures the percentage of responses that are valid, parseable JSON objects. It additionally performs validation against the ASN.1 specification of LDAP, following the same approach as in the parser. In practice, it penalizes any output that is truncated, contains unmatched brackets, or has incorrect types that prevent it from being interpreted as an LDAPMessage.

\item \textbf{Structure Pass Rate}: This metric represents the percentage of responses 
that contain the expected operation according to the request performed, applying strict 
rules. For instance, a \textit{bindRequest} must be followed only by a \textit{bindResponse}; a \textit{searchRequest} must include one or more \textit{searchResEntry} messages followed by a closing \textit{searchResDone}; modification operations *modify, add, del, modDN) must contain their respective responses; and an \textit{abandonRequest} must not produce any response. The goal is to verify that the minimal structure is respected with no additional or unintended operations.
    
\item \textbf{Key Field Accuracy}: This metric evaluates the consistency of key fields 
that must match between request and response. Mismatches between the values are strongly 
penalized. Additionally, it compares the expected and predicted operation types using a 
Jaccard  similarity index, as shown in Eq.~\ref{eq:jaccard}. The final score is the average of 
both components.

\begin{equation}
Jaccard(A,B) = \frac{|A \cap B|}{|A \cup B|}
\label{eq:jaccard}
\end{equation}

\[
A = \text{output\_protocolOps}, \quad 
B = \text{predicted\_protocolOps}
\]

\item \textbf{Completeness Score}: This metric assesses the completeness of responses, 
applied exclusively to \textit{searchRequest} operations. It considers two factors: the 
coverage of returned entries and the presence of a closing \textit{searchResDone}. Coverage is 
measured by comparing the number of predicted \textit{searchResEntry} elements with those 
expected; the expected value is taken from the reference output, and if unavailable, the 
sizeLimit from the request or at least one entry is used as a baseline. For non-search 
operations, the metric returns NaN; in such cases, its weight is automatically 
redistributed to the Structure Pass Rate to avoid artificially inflating the overall 
score.

\item \textbf{Weighted Validity Score}: This metric aggregates all previous ones into a single score. Two 
formulas are applied depending on the type of operation: one for searches (where completeness is explicitly 
considered), and another for non-search operations (where the weight of completeness is redistributed to 
structure). The specific formulations are illustrated in Eq.~\ref{eq:weighted_score_1} and Eq.~\ref{eq:weighted_score_2}. In both cases, syntactic robustness and structural validity are prioritized over key field accuracy or completeness, reflecting the principle that the most important requirement is for responses to be well-formed and contain the minimum required operation. A further advantage of this approach is its flexibility: weights can be adjusted according to system needs, either favoring robustness against malformed outputs or emphasizing accuracy and completeness, thus providing an adaptable evaluation metric.
\end{itemize}

\[
\text{Search Operation Formula}\quad 
\]
\begin{align}
\mathit{Weighted} = & \; 0.4 \cdot \mathit{Syntax} 
+ 0.3 \cdot \mathit{Structure} \nonumber \\
& + 0.2 \cdot \mathit{KeyFields} 
+ 0.1 \cdot \mathit{Completeness}
\label{eq:weighted_score_1}
\end{align}

\[
\text{Non-Search Operation Formula}\quad 
\]
\begin{align}
\mathit{Weighted} = & \; 0.4 \cdot \mathit{Syntax} 
+ 0.4 \cdot \mathit{Structure} \nonumber \\
& + 0.3 \cdot \mathit{KeyFields} 
\label{eq:weighted_score_2}
\end{align}

Based on this system, the evaluation script supports two execution modes.

In the first mode (\textit{--infer} option), the script takes as input a CSV file with two 
columns (input and output), where input corresponds to the LDAP request in JSON format 
and output represents the expected correct response. From this file, the script invokes 
the Bridge LLM functionality to connect to Colab and submit the requests. The 
predictions generated by the model are then stored alongside the request–response pairs, 
producing a new CSV file with three fields: input, output, and prediction. This file is 
subsequently processed by the evaluation system described earlier, generating a detailed 
report for each evaluated row as well as an aggregated summary with average metrics.

In the second mode (\textit{--data} option), the script starts from a CSV file that already 
contains all three fields (input, output, prediction). In this case, the inference phase 
is skipped, and the system proceeds directly to evaluation. This mode enables faster 
refinement and validation of the metrics, avoiding the need for new LLM calls and 
thereby accelerating testing cycles.

\begin{figure}[htbp]
\centering
\begin{tikzpicture}
\pie[
    text=legend,
    radius=2,
    sum=auto,
    color={blue!40, red!40, green!40, orange!40}
]{
44.62/authentications (44.62\%),
35.38/queries (35.38\%),
13.86/others (13.86\%),
6.15/modifications (6.15\%)
}
\end{tikzpicture}
\caption{Distribution of LDAP Operations in the Evaluation Dataset.}
\label{fig:ldap_operations_distribution_simplified}
\end{figure}
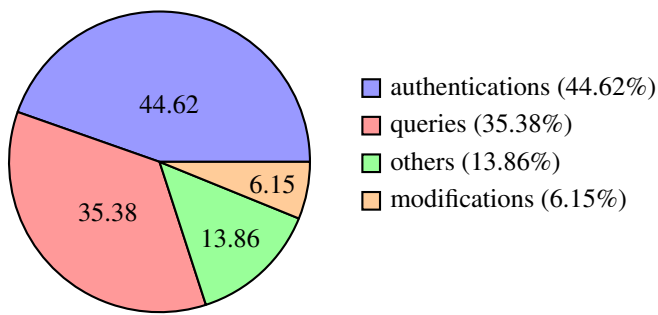

For the evaluation dataset, a broad range of LDAP operations was incorporated, 
including authentications through \textit{bindRequest} (44.62\%), queries via \textit{searchRequest} 
(35.38\%), modifications (6.15\%), comparisons (4.62\%), deletions (3.08\%), additions 
(3.08\%), and extended operations (3.08\%), the distribution can be observed in Fig.~\ref{fig:ldap_operations_distribution_simplified}. In total, the dataset comprises 65 records, 
with failed operations also included in order to replicate more realistic conditions. 
In this case, the domain used was uc3m.es, which had not been shown to the LLM during 
the training phase, thereby ensuring the independence of the evaluation data from the 
training set. Moreover, the use of this domain also makes it possible to observe how 
effectively the model is able to infer context in scenarios not explicitly covered 
during training.

The dataset was generated following the same collection process described previously: 
executing operations with different tools, capturing the traffic with Wireshark, and 
then processing the resulting files with the same custom script to produce a unified CSV.

\subsection{Results \& Metrics}

This section presents the results obtained from testing and evaluating the proposed 
system. The experiments were designed to validate both the functional behavior of the 
honeypot and its ability to correctly simulate LDAP interactions using an LLM.

It is important to note that all evaluations were performed under the same conditions as 
the training phase, using the same hardware configuration and the same LLM model. This 
ensures consistency across the experimentation pipeline and eliminates variability 
derived from changes in computational resources.

As a first step, individual tests were carried out using small scripts that executed a 
sequence of LDAP operations: a \textit{bindRequest}, a \textit{searchRequest} configured to produce 
multiple responses, and finally an \textit{unbindRequest}. The traffic generated during these 
operations was captured with Wireshark and inspected at the network level. The analysis 
confirmed that the system functioned correctly, with well-formed TCP segments, flags, 
and acknowledgments. At the application level, the LDAP responses were also verified to 
be correct, ensuring proper end-to-end behavior of the system.

Another simple but relevant experiment was conducted using the graphical application 
Apache Directory Studio. A search operation limited to five users was performed, 
requesting multiple attributes. The test was successful, and the results were displayed 
in the application’s interface as expected. This experiment not only confirmed that the 
honeypot can respond adequately to a real LDAP client application, but also demonstrated 
that the generated messages were syntactically correct and structurally consistent, 
enabling them to be parsed and visualized without errors.

These initial validation tests confirm that the system behaves correctly at both the 
network and application levels, providing a solid foundation before moving on to more 
systematic evaluations with the constructed datasets.

Shifting from functional validation to evaluation with the constructed dataset, the 
first step was to measure how well the model performed without any fine-tuning. This 
baseline evaluation provided an essential reference point, making it possible to later 
contrast the results against those obtained after fine-tuning. By establishing this 
comparison, it becomes clearer to what extent fine-tuning improves the system’s ability 
to generate robust and accurate LDAP responses under realistic conditions.

The results of the baseline evaluation are summarized in Table \ref{tab:evaluation_metrics_results}. 
Overall, the metrics reflect that, while the model is able to generate syntactically valid outputs in most 
cases, significant limitations remain in terms of structural consistency and 
completeness.

\renewcommand{\arraystretch}{1.5}
\begin{table}[htbp]
\caption{Baseline Model Evaluation}
\begin{center}
\footnotesize
\begin{tabular}{|m{3.5cm}|m{2.5cm}|}
\hline
\textbf{Metric} & \textbf{Value} \\
\hline
Syntax Pass Rate & 0.9231 \\
\hline
Structure Pass Rate & 0.5077 \\
\hline
Key Field Accuracy & 0.4723 \\
\hline
Completeness Score & 0.1000 \\
\hline
Weighted Validity Score & 0.6618 \\
\hline
\end{tabular}
\label{tab:evaluation_metrics_results}
\end{center}
\end{table}
\renewcommand{\arraystretch}{1.0}

The Syntax Pass Rate exceeded 90\%, which can be partly attributed to the fact that JSON 
is a widely adopted format and that the prompts were explicitly designed to reinforce 
correct syntax. By contrast, the Structure Pass Rate, just above 50\%, reveals that the 
baseline model frequently failed to correctly associate requests with their expected 
responses, reflecting that without training it lacks an understanding of which operation 
pairs belong together. Regarding Key Field Accuracy, which averaged around 47\%, the 
analysis shows that most of the penalties came from the Jaccard similarity on the 
operation type rather than from \textit{messageID} mismatches, suggesting that while identifiers 
were often preserved, the predicted operation class was not always aligned with the 
expected one. The Completeness Score, close to 10\%, further confirms that the baseline 
model struggled in particular with search operations, where coverage and proper 
termination were rarely achieved.

The Weighted Validity Score places overall performance at around two-thirds of the 
maximum possible value. This confirms that the baseline model is capable of producing 
syntactically correct outputs but lacks a reliable grasp of LDAP semantics, particularly 
in terms of operation pairing and search completeness. These limitations make clear the 
necessity of fine-tuning to strengthen structural validity and improve overall protocol 
fidelity.

With regard to the evaluation of the fine-tuned model, it should be mentioned that a 
small number of preliminary experiments were carried out to adjust training parameters 
such as the number of epochs and additional configuration details. These adjustments 
were only intended to reach a reasonable balance between accuracy and coherence in the 
generated responses. The results of these trials are not included here, as the primary 
goal of this work is not to optimize the model, but rather to demonstrate that the 
honeypot is capable of responding adequately to LDAP requests, as discussed in the introduction regarding its scope. 

The results of the fine-tuned model are summarized in Table \ref{tab:evaluation_metrics_results_2}. 
Compared with the baseline, the improvements are substantial across all metrics, demonstrating that fine-
tuning was critical to achieve robust LDAP simulation.

\renewcommand{\arraystretch}{1.5}
\begin{table}[htbp]
\caption{Fine-Tuned Model Evaluation}
\begin{center}
\footnotesize
\begin{tabular}{|m{3.5cm}|m{2.5cm}|}
\hline
\textbf{Metric} & \textbf{Value} \\
\hline
Syntax Pass Rate & 1.0000 \\
\hline
Structure Pass Rate & 1.0000 \\
\hline
Key Field Accuracy & 0.9769 \\
\hline
Completeness Score & 0.8132 \\
\hline
Weighted Validity Score & 0.9888 \\
\hline
\end{tabular}
\label{tab:evaluation_metrics_results_2}
\end{center}
\end{table}
\renewcommand{\arraystretch}{1.0}

The Syntax Pass Rate reached 100\%, confirming that every response generated was a valid 
and parseable JSON object. While the baseline already performed relatively well in this 
metric, the fine-tuned model completely eliminated cases of malformed or truncated 
outputs. This shows that the model internalized the reinforced syntax rules applied 
during training, guaranteeing syntactic reliability regardless of the complexity of the 
request.

The Structure Pass Rate also attained 100\%, a striking contrast with the baseline’s 
performance just above 50\%. This result resolves one of the baseline model’s main 
shortcomings, as it consistently demonstrates the ability to reproduce LDAP’s 
operational logic and correctly pair requests with their corresponding responses. This 
improvement directly addresses one of the most critical limitations of the baseline 
model, namely its inability to reproduce the operational logic of LDAP without 
supervision.

The Key Field Accuracy improved to nearly 98\%. A detailed review shows that all 
\textit{messageID} values were correctly preserved, and the remaining penalties in this metric 
were exclusively due to the Jaccard similarity on the operation type. The fine-tuned 
model not only preserved identifiers consistently but also learned to map the correct 
class of operations, thereby ensuring coherence between client requests and honeypot 
responses.

The Completeness Score rose to above 80\%, which is remarkable given that this metric 
was the weakest point of the baseline model at only 10\%. This shows that the fine-tuned 
model is capable of producing search results that closely match the expected coverage, 
while also ensuring the presence of the required closing \textit{searchResDone}. Since search 
operations dominate typical LDAP traffic, this improvement is critical to making the 
honeypot convincing in realistic attack scenarios.

Finally, the Weighted Validity Score reached almost 99\%, which reflects a near-optimal 
balance across syntax, structure, key fields, and completeness. Unlike the baseline 
model, whose weighted score highlighted inconsistencies in semantic validity, the fine-
tuned model demonstrates that it can deliver outputs that are not only syntactically 
correct but also semantically faithful to the LDAP specification.

In summary, fine-tuning transformed the model from one that could only approximate LDAP 
responses into a system capable of reliably simulating the protocol with high fidelity. 
The improvements in structural validity and completeness are particularly important, as 
they determine whether the honeypot is perceived as realistic by potential attackers. 
These results validate the fine-tuning approach as essential for bridging the gap 
between syntactic correctness and full protocol compliance.

The results show that the model is highly robust in simple LDAP operations such as 
\textit{bindRequest}, \textit{addRequest}, \textit{delRequest}, \textit{compareRequest}, and \textit{extendedReq}, where it achieved 
100\% in syntax, structure, and key fields, resulting in perfect overall scores. This 
indicates that in basic interaction scenarios the system is stable and reliable, 
accurately reproducing the expected behavior.

In contrast, \textit{searchRequest} operations represent the main weakness. Although syntax and 
structure are always correct, there is a noticeable drop in key fields (93.5\%) and, 
more significantly, in completeness (81.3\%). This means that the model tends to return 
partial responses, omitting some \textit{searchResEntry} elements before closing with 
\textit{searchResDone}. As a result, overall performance is slightly affected in more complex 
scenarios involving larger result sets, suggesting that further improvements are needed 
in handling long responses.

For further detail, Table \ref{tab:metrics_by_operation} presents the disaggregated results by 
operation type. As shown, the model consistently achieves perfect scores in 
simple operations such as \textit{bindRequest}, \textit{addRequest}, \textit{delRequest}, \textit{compareRequest}, and \textit{extendedReq}. The performance of \textit{searchRequest} operations, however, is comparatively
lower in terms of key fields and completeness, which confirms the observations discussed above.

\renewcommand{\arraystretch}{1.5}
\begin{table*}[htbp]
\caption{Evaluation Metrics by Operation}
\begin{center}
\footnotesize
\begin{tabular}{|m{2.5cm}|m{1cm}|m{1.6cm}|m{2cm}|m{1.8cm}|m{2cm}|m{2.2cm}|}
\hline
\textbf{Operation} & \textbf{Cases} & \textbf{Syntax Pass} & \textbf{Structure Pass} & \textbf{Key Fields} & \textbf{Completeness} & \textbf{Weighted Score} \\
\hline
\textbf{bindRequest}    & 29 & 1.00 & 1.00 & 1.00  & –     & 1.00 \\
\hline
\textbf{addRequest}     & 2  & 1.00 & 1.00 & 1.00  & –     & 1.00 \\
\hline
\textbf{delRequest}     & 2  & 1.00 & 1.00 & 1.00  & –     & 1.00 \\
\hline
\textbf{compareRequest} & 3  & 1.00 & 1.00 & 1.00  & –     & 1.00 \\
\hline
\textbf{extendedReq}    & 2  & 1.00 & 1.00 & 1.00  & –     & 1.00 \\
\hline
\textbf{searchRequest}  & 23 & 1.00 & 1.00 & 0.935 & 0.813 & 0.968 \\
\hline
\end{tabular}
\label{tab:metrics_by_operation}
\end{center}
\end{table*}
\renewcommand{\arraystretch}{1.0}

In this sense, the experiments confirm that the approach is feasible and that the tool fulfills its intended purpose. These findings provide the basis for the discussion 
presented in the following section, where the main conclusions of the work are outlined 
together with its limitations and potential directions for future research.

\section{Conclusion and Future Work}

This work has presented the design and implementation of a functional honeypot based on LLMs, 
capable of simulating the LDAP protocol with reasonable reliability and robustness. The 
primary objective, developing a deception system that attracts attackers, records their 
interactions, and generates reliable LDAP responses, has been achieved. As with any honeypot, 
the system enables the extraction of IOCs and TTPs from the interactions it records. These 
records can also be integrated into monitoring environments, such as a SIEM, where any 
interaction with the honeypot itself may serve as an early warning of an attempted compromise.

The evaluation results demonstrate not only that the system is capable of producing adequate 
LDAP responses, but also that the decision to fine-tune the model was essential in achieving 
this outcome. Fine-tuning significantly improved performance across all metrics, with notable 
gains in syntax and structural validity. The model learned to consistently reproduce the 
expected request–response pairs, eliminating the baseline errors where operations were 
mismatched. Similarly, improvements in key fields confirmed that the model reliably preserved 
identifiers and maintained coherence across multiple operations. Perhaps most importantly, the 
handling of \textit{searchRequest} operations, which represent the majority of real LDAP traffic, 
showed substantial progress compared with the baseline, underscoring the practical relevance 
of the improvements.

Additional mechanisms implemented during development also contributed to the robustness of the 
system. Token management strategies enabled faster responses for short operations, addressing 
one of the main disadvantages of LLMs while allowing parameters to be tuned for operations 
that required longer outputs. The automatic discarding of malformed responses and the 
correction of incomplete search operations (e.g., appending a missing \textit{searchResDone}) further 
strengthened the reliability of the honeypot, ensuring that its behavior remained consistent 
even in edge cases.

The custom evaluation framework proved to be well-suited to the specific requirements of LDAP, 
avoiding reliance on traditional machine learning metrics such as accuracy or F1-score, which 
would not provide meaningful insights in this context. Instead, the system evaluated responses 
according to protocol semantics and structural correctness. This design not only supports the 
assessment of the current implementation but also allows future extensions, for instance by 
placing greater emphasis on completeness or incorporating new evaluation dimensions.

Taken together, the integration of an LLM at the core of the honeypot, the supporting 
mechanisms for token management and response validation, and the custom evaluation system 
provide a flexible framework that can be adapted to different scenarios and improved over 
time. The work therefore demonstrates both the feasibility of using LLMs for deception 
purposes and the potential for refinement of the proposed system.

Beyond the tool itself, this work addresses a clear gap in the literature. To the best 
of our knowledge, no previous research has proposed or implemented an LLM-based 
honeypot for LDAP with comparable depth. The contribution is therefore twofold: on the 
one hand, a novel honeypot system specifically tailored to LDAP available on Github \cite{llm_ldap_honeypot}; and on the other, the 
creation of a dataset of LDAP traffic that is being made publicly available via 
platforms such as Hugging Face \cite{ldaptraffic_dataset}. This dataset constitutes a valuable 
resource for the community, enabling further research and experimentation in an area 
where publicly available data has been lacking.

While the results achieved demonstrate the feasibility and robustness of the proposed 
system, several aspects remain open for improvement and further exploration. Future 
work will focus on extending the current foundation with additional capabilities, 
enhancing realism, and addressing limitations that were intentionally left out of the 
present scope.

\begin{itemize}

\item \textbf{Encrypted LDAPS traffic}:
Although the parser is capable of encoding and decoding ASN.1 without crashing, the 
values remain encrypted when LDAPS is used. While much of the literature reports the 
presence of plaintext LDAP servers, in more mature environments it would be unusual to 
find an LDAP service without encryption, which could raise suspicion for an attacker. 
Furthermore, attempts to interact with the honeypot over LDAPS would likely produce 
unfavorable results. A possible solution would be to introduce an intermediate layer 
that handles TLS, ensuring that the LLM receives the decrypted traffic as plaintext 
while preserving the realism of encrypted communication.

\item \textbf{Exploring alternative models}:
In this project, LLaMA was chosen as the main model due to its strong performance and 
familiarity within the research community. However, one of the clearest avenues for 
improvement is to evaluate different LLMs, including lighter models with fewer 
parameters that could provide more real-time responses. An interesting candidate for 
future testing is Qwen2.5-7B \cite{qwen25-7b-instruct}, which in addition to being 
smaller, has been specifically designed to generate structured outputs such as JSON, 
making it particularly relevant for this use case.

\item \textbf{Expanding the dataset}:
Although the current dataset includes a wide range of operations and covers most of the 
possibilities defined in the LDAP protocol, its scope could be further extended. Future 
work could focus on enriching the dataset with more diverse operations, edge cases, and 
real-world interaction patterns observed in production systems. Increasing dataset 
variety would not only improve the training process but also enable the model to better 
handle unexpected or less frequent queries.

\item \textbf{Maintaining session context}:
One limitation of the current implementation is the lack of session persistence, which 
can result in inconsistencies. For example, repeating the same query may produce 
different results, a behavior that would likely raise suspicion for an attacker. To 
address this, future versions of the honeypot should incorporate mechanisms to maintain 
session context, storing previous requests and responses so that subsequent 
interactions remain coherent. Preserving this continuity would increase the realism of 
the honeypot, making it more convincing in long-lived attacker sessions and providing 
richer intelligence on adversarial behavior.

\end{itemize}

\section*{Acknowledgments}

Florina Almenares would like to acknowledge the support of R\&D projects: PID2023-148716OB-C32 (DISCOVERY) funded by the Spanish Ministry of Science, Innovation and Universities MICIU/AEI/10.13039/501100011033, PRTR-INCIBE-2023/00623/001 (I-Shaper Strategic Project), funded by the “European Union NextGenerationEU/PRTR” due to the collaboration agreement signed between the Instituto Nacional de Ciberseguridad (INCIBE) and the UC3M, and TEC-2024/COM-504 (RAMONES-CM), funded by the Comunidad de Madrid.

Alfonso Sánchez-Macián would like to acknowledge the support of R\&D project PID2022-136684OB-C21 (Fun4Date) funded
by the Spanish Ministry of Science and Innovation MCIN/AEI/10.13039/501100011033 and TUCAN6-CM (TEC-
2024/COM-460), funded by Comunidad de Madrid (ORDEN 5696/2024)

\bibliographystyle{IEEEtran}
\bibliography{references}

\end{document}